\documentclass[10pt,conference]{IEEEtran}
\IEEEoverridecommandlockouts
\normalsize
\usepackage{fixltx2e}
\usepackage{mathrsfs}
\usepackage{amsmath}
\usepackage{mathtools}
\usepackage{graphics}
\usepackage{float}
\usepackage{epstopdf}
\usepackage{graphicx}
\usepackage{bm}
\usepackage{amsthm}
\usepackage{tikz}
\usepackage[lined,boxed,commentsnumbered,linesnumbered, ruled]{algorithm2e}
\usepackage{comment}
\usepackage{balance}
\usepackage{cite}
\usepackage[capitalize]{cleveref}
\crefname{equation}{\unskip}{\unskip}
\usetikzlibrary{shapes, patterns,decorations.text, decorations.pathreplacing}
\usepackage{ifthen}

\newcommand*{\Scale}[2][4]{\scalebox{#1}{\ensuremath{#2}}}%

\newtheorem{example}{Example}
\newtheorem{theorem}{Theorem}
\newtheorem{definition}{Definition}
\newtheorem{lemma}{Lemma}
\newtheorem{corollary}{Corollary}

\newcommand{\creflastconjunction}{, and\nobreakspace}

\newcommand{\X}{\bm X}
\newcommand{\BDelta}{\boldsymbol{\Delta}}

\begin{document}

\title{Private Information Retrieval in Distributed Storage Systems Using an Arbitrary Linear Code} %
\author{\IEEEauthorblockN{Siddhartha Kumar\IEEEauthorrefmark{2}, Eirik Rosnes\IEEEauthorrefmark{2}, and Alexandre Graell i Amat\IEEEauthorrefmark{3}}\\
\vspace{-4mm}\IEEEauthorblockA{\IEEEauthorrefmark{2}Simula@UiB, N-5020 Bergen, Norway}
\IEEEauthorblockA{\IEEEauthorrefmark{3}Department of Electrical Engineering, Chalmers University of Technology, SE-41296 Gothenburg, Sweden
}
\thanks{The work of S. Kumar and E.\ Rosnes was partially funded by the Research Council of Norway (grant 240985/F20). A.\ Graell i Amat was supported by the Swedish Research Council under grant \#2016-04253.}}%

\maketitle

\begin{abstract}
We propose an information-theoretic private information retrieval (PIR) scheme for distributed storage systems where data is stored using a linear systematic  code of rate $R > 1/2$. The proposed scheme generalizes the PIR scheme for data stored using maximum distance separable codes recently proposed by Tajeddine and El Rouayheb for the scenario of a single spy node. We further propose an algorithm to optimize the communication price of privacy (cPoP) using the structure of the underlying linear code. As an example, we apply the proposed algorithm to several distributed storage codes, showing that the cPoP can be significantly reduced by exploiting the structure of the distributed storage code.
\end{abstract}

\section{Introduction}

In data storage applications, besides resilience against disk failures and data protection against illegitimate users, the privacy of the data retrieval query may also be of concern. For instance, one may be interested in designing a storage system in which a file can be downloaded without revealing any information of which file is actually downloaded to the servers storing it. This form of privacy is usually referred to as \emph{private information retrieval} (PIR). PIR is important to, e.g.,  protect users from surveillance and monitoring. 

PIR protocols were first studied by Chor \emph{et al.} in \cite{cho95}, which introduced the concept of an $n$-server PIR protocol, where a binary database is replicated among $n$ non-colluding servers (referred to as nodes) and the aim is to privately retrieve a single bit from the database while minimizing the total upload and download communication cost. The communication cost in \cite{cho95} was further reduced in \cite{Yek08} and references therein.  
  Since then, coded PIR schemes have been introduced, where the database is encoded (as opposed to simply being replicated) across several nodes \cite{ish04}. With the advent of distributed storage systems (DSSs), where the database is encoded and then stored on $n$ nodes, there has been an increasing interest in implementing coded PIR protocols for these systems. PIR protocols for DSSs, where data is stored using codes from two explicit linear code constructions (one protocol for each code construction), were presented in \cite{sha14}, 
%
 and information-theoretic lower bounds on the tradeoff between the storage cost and the retrieval cost were provided in \cite{cha15}. 
 In \cite{Faz15}, the authors introduced PIR codes which when used in conjunction with traditional $n$-server PIR protocols allow to achieve PIR on DSSs. These codes achieve high code rates without sacrificing on the communication cost of an $n$-server PIR protocol. Recently, the authors in \cite{taj16} proposed a coded PIR protocol for DSSs that use an $(n,k)$ maximum distance separable (MDS) code for storing data on $n$ storage nodes. The proposed protocol achieves privacy in the presence of at most $n-k$ colluding nodes. In addition, when there are no colluding nodes, the protocol achieves the lowest possible amount of downloaded data per unit of stored data, referred to as the \emph{communication price of privacy} (cPoP).

In the storage community, it is well known that MDS codes are inefficient in the repair of failed nodes. Repair is essential to maintain the initial state of reliability of the DSS. To address efficient repair, Pyramid codes \cite{Hua07} and locally repairable codes (LRCs) \cite{Sat13}, have been proposed. They achieve low locality, i.e., a low number of nodes need to be contacted to repair a single failed node. 
In this paper, for the scenario with no colluding nodes (i.e., a single spy node), we extend the PIR protocol from \cite{taj16} to a more general case where data is stored using an arbitrary systematic linear storage code of rate $R > 1/2$. 
We show that the cPoP can be optimized using the structure of the code, and we provide an algorithm to search for an optimal (in terms of the lowest possible cPoP) protocol. We present the optimal cPoP that can be achieved for various linear codes, including LRCs and Pyramid codes. 
Interestingly, our numerical results show that non-MDS codes can also achieve the lower bound on the cPoP provided in \cite{cha15}.
Our work bears some similarities to the parallel work in \cite{Hol16}, where a PIR protocol protecting against multiple colluding nodes for any linear storage code was presented. However, we show that our extended protocol achieves  better cPoP for the scenario of a single spy node.


\section{System Model}

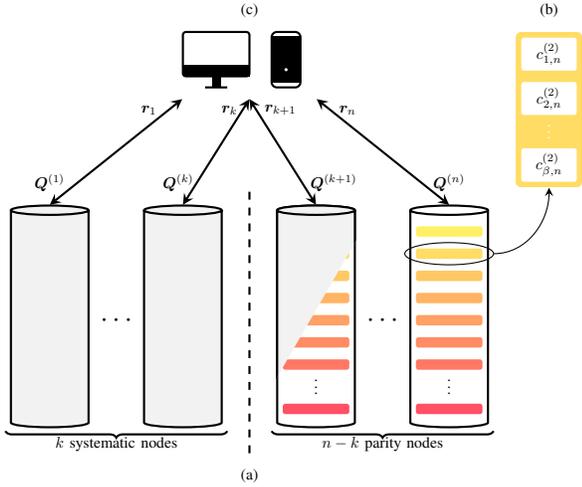
\begin{figure}[t]
\centering
\begin{tikzpicture}[scale=0.59, every node/.style={transform shape}, thick]

\node[minimum width=3cm, minimum height=2cm] (user) at (5.5,10.5) {};
\foreach \i in {1,4,7,10}{
\ifthenelse{\equal{\i}{1}\OR\equal{\i}{4}}
{\node[draw, black, fill=gray!10, cylinder, thick, aspect=0.9,minimum height=5 cm, minimum width=1.75 cm, rotate=90] (n\i) at (\i,4.5) {};}
{\node[draw, black, fill=white, cylinder, thick, aspect=0.9,minimum height=5 cm, minimum width=1.75 cm, rotate=90] (n\i) at (\i,4.5) {};}
}
\draw[dashed] (5.5,1.5)--(5.5,7.5);
\node at (2.55,4.5) {\LARGE $\cdots$};
\node at (8.55,4.5) {\LARGE $\cdots$};
\node at (10,3.1) {$\vdots$};
\node at (7,3.1) {$\vdots$};
\foreach \i in {2,3,3.5,4,4.5,5,5.5,6}{
	\pgfmathsetmacro\Gval{\i*40}
	\definecolor{mycolor\i}{RGB}{255,\Gval,100}
	\node[fill=mycolor\i, minimum height=0.15cm, minimum width=1.5cm, rounded corners=1pt] at (10,\i+0.5) {};
	\node[fill=mycolor\i, minimum height=0.15cm, minimum width=1.5cm, rounded corners=1pt] at (7,\i+0.5) {};
}
\definecolor{file}{RGB}{255,220,100}
\node[fill=file, rounded corners=2pt, minimum width=1.5cm, minimum height=3.5cm] (f) at (12.25,9.25) {};
\foreach \i/\j in {7.5/3,9/2,10/1}{
	\ifthenelse{\equal{\j}{3}}
	{\node[fill=white, minimum height=0.7cm, minimum width=1.25cm, rounded corners=1pt] at (12.25,\i+0.5) {$c^{(2)}_{\beta, n}$};}
	{\node[fill=white, minimum height=0.7cm, minimum width=1.25cm, rounded corners=1pt] at (12.25,\i+0.5) {$c^{(2)}_{\j, n}$};}
}
\node at (12.25,8.85) {\color{white}{$\vdots$}};
\node[draw, thin, ellipse, minimum width=2cm, minimum height=0.5cm] (ell) at (10,6) {};
\draw[>=stealth,->,thin] (ell.east) to[in=270,out=0] (f.south);
\begin{scope}[yshift=0cm]
	
\clip[shift={(-1,0.5)},rotate around={-30:(7,4.5)}] (6,2)rectangle(8,7.5);
\node[draw, black,fill=gray!10, cylinder, thick, aspect=0.9,minimum height=5 cm, minimum width=1.75 cm, rotate=90] at (7,4.5) {};	
\end{scope}
\begin{scope}[shift={(4,9.5)}]
	\draw[rounded corners=1.2pt] (0,0.5)rectangle(1.5,1.5);
	\path[fill=black] (0,0.5)rectangle(1.5,0.75);
	\draw[rounded corners=2pt] (2,0.25)rectangle(2.65,1.5);
	\path[fill=black] (2,0.35)rectangle(2.65,1.4);
	\path[fill=white] (2.325,0.65) circle (1pt);
	\path[fill=black] (0.65,0.25)rectangle(0.85,0.5);
	\draw[thick] (0.5,0.25)--(1,0.25);
\end{scope}

\draw[>=stealth,<->] (n1.east) -- (user.south west);
\draw[>=stealth,<->] (n4.east) -- (user.south);
\draw[>=stealth,<->] (n7.east) -- (user.south);
\draw[>=stealth,<->] (n10.east) -- (user.south east);
\node[shift={(0,0.5)}] at (n1.east) {$\bm Q^{(1)}$};
\node[shift={(-0.1,0.5)}] at (n4.east) {$\bm Q^{(k)}$};
\node[shift={(0.4,0.5)}] at (n7.east) {$\bm Q^{(k+1)}$};
\node[shift={(0,0.5)}] at (n10.east) {$\bm Q^{(n)}$};

\node[shift={(-0.75,-0.25)}] at (user.south west) {$\bm r_1$};
\node[shift={(-0.45,-0.25)}] at (user.south) {$\bm r_k$};
\node[shift={(0.7,-0.25)}] at (user.south) {$\bm r_{k+1}$};
\node[shift={(0.7,-0.25)}] at (user.south east) {$\bm r_n$};

\draw[decorate,decoration={brace,mirror}, thick] (0,2)--node[below]{$k$ systematic nodes}(5,2);
\draw[decorate,decoration={brace,mirror}, thick] (6,2)--node[below]{$n-k$ parity nodes}(11,2);

\node at (5.5,1) {(a)};
\node at (12.25,11.5) {(b)};
\node at (5.5,11.5) {(c)};

\end{tikzpicture}
\vspace{-2ex}
\caption{System Model. (a) The colored boxes in each storage node represent the $f$ coded chunks pertaining to the $f$ files. (b) Coded chunk corresponding to the $2$nd file in the $n$-th node. It consists of $\beta$ code symbols, $c^{(2)}_{i,n}, i=1,\ldots,\beta$. (c) The user sends the queries $\bm Q^{(j)},j=1,\ldots,n$, to the storage nodes and receives responses $\bm r_j$.}
\label{Fig: System Model}
\vskip -3ex
\end{figure}

%

We consider a DSS that stores $f$ files $\bm X^{(1)}, \bm X^{(2)},\ldots, \bm X^{(f)}$, where each file $\bm{X}^{(m)}=[x_{ij}^{(m)}]$, 
$m=1,\ldots,f$, is a $\beta \times k$ matrix over $\text{GF}(q^{\alpha \ell})$, with  $\beta$, $k$, $\alpha$, and $\ell$ being positive integers and 
%
$q$ some prime number. Each file is divided into $\beta$ stripes (blocks) and encoded using a linear code as follows. Let $\bm x^{(m)}_i=(x^{(m)}_{i,1},x^{(m)}_{i,2},\ldots,x^{(m)}_{i,k})$, $i=1,\ldots,\beta$, be a message vector (corresponding to the $i$-th row of $\X^{(m)}$) that is encoded by an $(n,k)$ linear code $\mathcal{C}$ over $\text{GF}(q^\alpha)$, having \emph{subpacketization} $\alpha$, into a length-$n$ codeword $\bm c^{(m)}_i=(c^{(m)}_{i,1},c^{(m)}_{i,2},\ldots,c^{(m)}_{i,n})$, where $c_{i,j}^{(m)}\in\text{GF}(q^{\alpha \ell})$.  
When $\alpha=1$, the code $\mathcal{C}$ is referred to as a \emph{scalar} code. Otherwise, the code is called a \emph{vector} code \cite{Bla95}. 
The $\beta f$ generated codewords $\bm c_i^{(m)}$ are then arranged in the array  $\bm C=((\bm c^{(1)}_1)^\top|\ldots|(\bm c^{(1)}_{\beta})^\top|(\bm c^{(2)}_1)^\top|\ldots|(\bm c^{(f)}_{\beta})^\top)^\top$ of dimension $\beta f  \times n$, where $(\cdot)^\top$ denotes the transpose of its argument and $(\bm v_1|\ldots|\bm v_{\beta f})$ denotes the concatenation of column vectors $\bm v_1,\ldots,\bm v_{\beta f}$. For a given column $j$ of $\bm C$, we denote the vector $(c_{1,j}^{(m)},c_{2,j}^{(m)},\ldots,c_{\beta,j}^{(m)})$ as a coded chunk pertaining to file $m$. Then the $f$ coded chunks in column $j$ are stored on the $j$-th node as shown in \cref{Fig: System Model}(a).
We also assume that the $(n,k)$ code $\mathcal C$ is systematic and that the first $k$ code symbols of $\bm c_i^{(m)}$ are message symbols. 
Accordingly, we say that the first $k$ nodes are systematic nodes and the remaining nodes are parity nodes.

\subsection{Privacy Model} \label{sec:privacy}
We consider a DSS where  any single node may act as a spy node. Let $s\in\{1,\ldots,n\}$ denote the spy node in the DSS. The role of the spy node is to determine which file, $m$, is accessed by the user. We assume that the user does not know $s$, since otherwise it can trivially achieve PIR by avoiding contacting the spy node.  In addition, the remaining non-spy nodes do not collaborate with the spy node. To retrieve  file $\bm X^{(m)}$ from the DSS, the user sends a $d\times\beta f$  matrix query $\bm Q^{(j)}=[q_{il}^{(j)}]$ over $ \text{GF}(q^\alpha) \subseteq \text{GF}(q^{\alpha \ell})$ to the $j$-th node for all $j \in \{1,\ldots,n\}$. 
Depending on the queries, node $j$ sends the  column vector
\begin{equation} \label{eq:response}
	\bm r_j=(r_{j,1},\dots,r_{j,d})^\top = \bm Q^{(j)}(c^{(1)}_{1,j},c^{(1)}_{2,j},\ldots,c^{(1)}_{\beta,j},\ldots,c^{(f)}_{\beta,j})^\top,
\end{equation}
referred to as the response vector, back to the user as illustrated in \cref{Fig: System Model}(c).  The following definition shows how such a scheme can achieve perfect information-theoretic PIR.
\begin{definition}
	Consider a DSS with $n$ nodes storing $f$ files in which a node $s\in\{1,\ldots,n\}$ acts as a spy. A user who wishes to retrieve the $m$-th file sends the queries $\bm Q^{(j)}$, $j=1,\ldots,n$, to all storage nodes, which return the responses $\bm r_j$. This scheme achieves perfect information-theoretic PIR if and only if
\begin{subequations} \label{Def:cond}
	\begin{align}
	\label{Def: cond1}
		{\rm Privacy:}\,\,\,\,	 &H(m|\bm Q^{(s)})=H(m)\\
	\label{Def: cond2}
		{\rm Recovery:}\,\,\,\, &H(\bm X^{(m)}|\bm r_1,\bm r_2,\ldots, \bm r_n)=0,
	\end{align}
\end{subequations}
where $H(\cdot)$ denotes the entropy function. 

\end{definition}
Queries satisfying \cref{Def: cond1} ensure that the spy node is not able to determine which file is being downloaded by the user. The recovery constraint in \cref{Def: cond2} ensures that the user is able to recover the requested file from the responses sent by the DSS. 

The efficiency of a PIR scheme is defined as the total amount of downloaded data per unit of retrieved data, since it is assumed that the content of the retrieved file dominates the total communication cost, i.e., $\ell$ is much larger than $f$ \cite{taj16}.

\begin{definition}
The cPoP of a PIR scheme, denoted by $\theta$,  is the total amount of downloaded data per unit of retrieved data,
\begin{align*}
	\theta=\frac{nd}{\beta k}.
\end{align*}
\end{definition}

It was shown in \cite[Th.~3]{cha15} that the cPoP for a DSS with a single spy node is lowerbounded by $\frac{1}{1-R}$ for a special kind of linear retrieval schemes, where $R$ is the rate of the linear code used to store the data in the DSS. In the case of more than one spy node, an explicit lower bound is currently unknown.

%

\section{Construction} \label{sec:constr}
In this section, we present a PIR scheme for a DSS where any node may be a spy node. 
The DSS uses an $(n,k)$ systematic linear  code over GF$(q^\alpha)$, of rate $R=k/n>1/2$ and subpacketization $\alpha$. The code is defined by its parity-check matrix, $\bm H$, of size $(n-k)\times n$, and its minimum distance is denoted by $d_{\text{min}}$. Since the code is systematic, $\bm H$ can be written as  $\bm H=(\bm P | \bm I)$, 
where $\bm I$ is an $(n-k)\times(n-k)$ identity matrix and $\bm P$ is an $(n-k)\times k$ parity matrix. In the following, let $\tilde{d}_\text{min}$ denote the minimum distance of the $(\tilde{n}=k,\tilde{k} \geq 2k-n)$ code, denoted by $\tilde{\mathcal{C}}$, defined by the parity-check matrix $\tilde{\bm H}=\bm P$. We choose $d=k$ and design the $n$ queries as
\begin{align}
\label{Eq: Query_design}
	\bm Q^{(l)}=\begin{cases}
		\bm U+\bm V^{(l)}, & \text{if $l=1,\ldots,k$}\\
		\bm U, & \text{if $l=k+1,\ldots,n$}		
	\end{cases},
\end{align}
where $\bm U=[u_{ij}]$ is a $k \times \beta f$ matrix whose elements $u_{ij}$ are chosen independently and uniformly at random from GF$(q^\alpha)$, and $\bm V^{(l)}=[v_{ij}^{(l)} ]$ is a $k\times\beta f$ deterministic binary matrix over $\text{GF}(q^\alpha)$.
Note that each $k\times \beta f$ query matrix $\bm Q^{(l)}$ represents $k$ subqueries, where each subquery corresponds to a row of $\bm Q^{(l)}$, 
and where $v^{(l)}_{ij}=1$ means that the $j$-th symbol in node $l$ is accessed by the $i$-th subquery of $\bm Q^{(l)}$.

Let $\bm E=[e_{ij}]$ be a $k\times k$ binary matrix, where $e_{ij}=1$ represents the $i$-th subquery of the $j$-th query accessing a message symbol. The design of $\bm V^{(l)}$ depends on the structure of $\bm E$, which must satisfy the following conditions.
\begin{list}{\labelitemi}{\leftmargin=1.5em}
\addtolength{\itemsep}{0.1\baselineskip}
	\item[1)] The user should be able to recover $\beta$ unique symbols of the requested file $\bm{X}^{(m)}$ from the $i$-th subquery of all $k$ queries, i.e., each row of $\bm E$ should have exactly $\beta$ ones. \label{item:1}
	\item[2)] The user should be able to recover $\beta$ unique symbols of the requested file $\bm X^{(m)}$ from each query (consisting of $k$ subqueries), i.e., each column in $\bm E$ should have $\beta$ ones. \label{item:2}
\item[3)] The user should be able to recover all $\beta k $ symbols of the requested file $\bm X^{(m)}$. This means that all rows of $\bm E$ (considered here as length-$k$ erasure patterns, with a one indicating an erasure) should be correctable by a maximum likelihood (ML) decoder for the $(\tilde{n}=k,\tilde{k} \geq 2k-n)$ code $\tilde{\mathcal{C}}$ on the binary erasure channel (BEC), i.e., the rows of $\bm E$ considered as erasure patterns are ML-correctable by $\tilde{\mathcal{C}}$. \label{item:3}
\end{list}

From conditions 1) and 2) it follows that $\bm E$ is a regular matrix with $\beta$ ones in each row and column. Condition 3) ensures the recovery condition (see (\ref{Def: cond2})).  Details are given in the proof of \cref{th:PIR}. Given $\bm E$, $\bm V^{(l)}$ has the following structure
\begin{align*}
	\bm V^{(l)}=\left(\begin{matrix}
	\bm 0_{k\times(m-1)\beta} & | & \Scale[1]{\BDelta}_l & | & \bm 0_{k\times(f-m)\beta} 
\end{matrix}\right),
\end{align*}
where $\bm 0_{i\times j}$ denotes the $i\times j$ all-zero matrix and $\BDelta_l$  is a $k\times\beta$ binary matrix. For $l=1,\ldots,k$, 
\begin{align} 
	\label{Eq: delta_design}
	\Scale[1]{\BDelta}_l= 
\left(\begin{matrix}
\bm \omega^\top_{\pi(j_1^{(l)})} & | &  \bm \omega^\top_{\pi(j_2^{(l)})} & | & \ldots & | & \bm \omega^\top_{\pi(j_k^{(l)})} 
	\end{matrix}\right)^\top,
\end{align}%
where $\pi:\{0,\ldots,\beta\}\rightarrow\{0,\ldots,\beta\}$ is an arbitrary permutation of size $\beta+1$ with a fixed point at zero, i.e., $\pi(0)=0$, $\bm \omega_t$, $t=1,\ldots,\beta$, is the $t$-th $\beta$-dimensional unit vector, i.e., a length-$\beta$ weight-$1$ binary vector with a single $1$ at the $t$-th position, $\bm \omega_0$ is the all-zero vector of length $\beta$, and 
\begin{align*}
	j_i^{(l)}=\begin{cases}
		{z}_{i}^{(l)}, & \text{if $e_{i l}=1$}\\
		0, & \text{otherwise}
	\end{cases},
\end{align*}
where $z_i^{(l)}\in\{1,\ldots,\beta\}$ and $z_i^{(l)}\not=z_{i'}^{(l)}$  for $i\not=i'$, $i,i'=1,\ldots,k$. In the following lemma, we show that such a construction of the queries ensures that the \emph{privacy} condition  \cref{Def: cond1} is satisfied.

\begin{lemma}
	\label{Lem: Privacy}
	Consider a DSS that uses an $(n,k)$ linear code with subpacketization $\alpha$ to store $f$ files, each divided into $\beta$ stripes, and assume the privacy model with a single spy node. Then, the queries $\bm Q^{(j)}$, $j=1,\ldots,n$, designed as in \cref{Eq: Query_design} satisfy $H(m|\bm Q^{(s)})=H(m)$, where $s\in\{1,\ldots,n\}$ is the spy node.
\end{lemma}

\begin{IEEEproof}
The queries $\bm Q^{(j)}$, $j=1,\ldots,k$, are a sum of a random matrix $\bm U$ and a deterministic matrix $\bm V^{(j)}$. The resulting queries have elements that are independently and uniformly distributed at random from GF$(q^\alpha)$. The same holds for the remaining queries as they are equal to $\bm U$. Hence, any $\bm Q^{(j)}$ obtained by the spy node is statistically independent of $m$. This ensures that $H(m|\bm Q^{(s)})=H(m)$.
\end{IEEEproof}

In order to show that the proposed PIR protocol achieves perfect information-theoretic PIR, it remains to be proved that from the responses $\bm r_j$ in \cref{eq:response}, sent by the nodes back to the user, one can recover the requested file, i.e., that the constructed PIR scheme satisfies the \emph{recovery} condition in \cref{Def: cond2}. We call each symbol of the response $\bm r_j$ a subresponse symbol generated from the corresponding subquery.


\begin{theorem} \label{th:PIR}
	Consider a DSS that uses an $(n,k)$ linear code with subpacketization $\alpha$ to store $f$ files, each divided into $\beta$ stripes. In order to retrieve the file $\bm X^{(m)}$, $m=1,\ldots,f$, from the DSS, the user sends the queries $\bm Q^{(j)}$, $j=1,\ldots,n$, in \cref{Eq: Query_design} to the $n$ storage nodes. Then, for the responses $\bm r_j$ in \cref{eq:response} received by the user, $H(\bm X^{(m)}|\bm r_1,\bm r_2,\ldots, \bm r_n)=0$.
\end{theorem}

\begin{IEEEproof}
See Appendix.
\end{IEEEproof}

\cref{th:PIR} generalizes \cite[Th.~1]{taj16} to any linear code with rate $R>1/2$. 
We remark that for the theorem to hold there is an implicit assumption that the three requirements for the matrix $\bm E$ mentioned above are all satisfied  ($\bm E$ is used in the construction of the queries $\bm Q^{(j)}$). Thus, the parameter $\beta$ (which is not explicitly mentioned in the theorem) has to be carefully selected such that a $\beta$-regular matrix $\bm E$ (satisfying the third requirement)  actually exists. In the following corollary, we provide such a particular value of $\beta$. 


\begin{corollary} \label{cor:PIR}
For $\beta=\tilde{d}_{\rm min}-1$, it holds that 
\begin{align}
\label{eq:Hth}
H(\bm X^{(m)}|\bm r_1,\bm r_2,\ldots, \bm r_n)=0,
\end{align}
and the cPoP becomes $\theta=\frac{n}{\beta}=\frac{n}{\tilde d_\text{min}-1}$.
\end{corollary}

\begin{IEEEproof}
\eqref{eq:Hth} follows directly from \cref{th:PIR}, since all erasure patterns of weight less than $\tilde{d}_{\text {min}}$ are ML-correctable, and $\theta=\frac{n}{\tilde d_\text{min}-1}$ follows also from $d=k$.
\end{IEEEproof}

In \cite{Hol16}, a PIR protocol achieving a cPoP of $\bar\theta\geq\frac{n}{d_\text{min}-1}$, with equality when $d_{\text{min}}-1$ is a divisor of $k$, 
was given. Note that $d_{\text{min}}\leq\tilde d_{\text{min}}$, and thus $\theta\leq\bar\theta$ for our construction.



\begin{example} \label{ex:5_3_code}
	Consider a DSS that uses a $(5,3)$ scalar ($\alpha=1$) binary linear code to store a single file by dividing it into $\beta$ stripes. 
	The code is defined by the  parity-check matrix
	\begin{align*}
		\bm H=(\bm P|\bm I)=\left(\begin{matrix}
			1 & 1 & 0 & 1 &0\\
			0 & 1 & 1 & 0 &1
		\end{matrix}\right).
	\end{align*}
	To determine the value of the parameter $\beta$, we compute the minimum distance $\tilde d_\text{min}$ of the $(\tilde n=3,\tilde k=1)$ linear code with parity-check matrix $\tilde{\bm H}=\bm P$. 
	From $\tilde{\bm H}$ it follows that $\tilde d_\text{min}=3$. Hence, from \cref{cor:PIR}, $\beta=\tilde d_\text{min}-1=2$. Let the file to be stored be denoted by the $2\times 3$ matrix $\bm X=[x_{ij}]$, where the message symbols $x_{ij}\in\text{GF}(2^\ell)$ for a positive integer $\ell$.  Then,
	\begin{align*}
		\bm C=\left(\begin{matrix}
			x_{11} & x_{12} & x_{13} & x_{11}+x_{12} & x_{12}+x_{13}\\
			x_{21} & x_{22} & x_{23} & x_{21}+x_{22} & x_{22}+x_{23}
		\end{matrix}\right).
	\end{align*}
	The user wants to download the file $\bm X$ from the DSS and sends a query $\bm Q^{(j)}$, $j=1,\ldots,5$, to the $j$-th storage node. The queries take the form shown in \cref{Eq: Query_design}. For $l=1,\ldots,3$, we construct the matrix $\bm V^{(l)}=\BDelta_l$ by choosing an appropriate $\bm E$. The only condition in the choice of $\bm E$ is that it is $\beta$-regular. We choose
	\begin{align*}
		\bm E=\left(\begin{matrix}
			1 & 0 & 1\\
			1 & 1 & 0\\
			0 & 1 & 1
		\end{matrix}\right)
	\end{align*}
	and construct $\BDelta_1$ according to \cref{Eq: delta_design}. Focusing on the first column of $\bm E$, we can see that the first two rows have a one in the first position. Thus, we choose $j_1^{(1)}=2$, $j_2^{(1)}=1$, and $j_3^{(1)}=0$, since $e_{11}=1$, $e_{21}=1$, and $e_{31}=0$, and take the permutation $0 \rightarrow 0$, $1 \rightarrow 2$, and $2 \rightarrow 1$ as $\pi$ to get 
	\begin{align*}
		\BDelta_1 =\left(\begin{matrix}
		\bm \omega_{\pi(2)}\\
		\bm \omega_{\pi(1)}\\
		\bm \omega_{\pi(0)}
		\end{matrix}\right)=\left(\begin{matrix}
		1 & 0\\
		0 & 1\\
		0 & 0
	\end{matrix}\right).
	\end{align*}
	In a similar fashion, we construct 
	\begin{align*}
		&\BDelta_2=\left(\begin{matrix}
			0&0\\
			1&0\\
			0&1
		\end{matrix}\right) \text{and}\;
		\BDelta_3=\left(\begin{matrix}
			0&1\\
			0&0\\
			1&0
		\end{matrix}\right).
	\end{align*}
%
%
	The queries $\bm Q^{(j)}$ are sent to the respective nodes and the responses
	\begin{align*}
           \begin{split}
		\bm r_1
                  &=\left(\begin{smallmatrix}
			u_{11}x_{11}+u_{12}x_{21}+x_{11}\\
			u_{21}x_{11}+u_{22}x_{21}+x_{21}\\
			u_{31}x_{11}+u_{32}x_{21}
		\end{smallmatrix}\right)=\left(\begin{smallmatrix}
			I_1+x_{11}\\
			I_4+x_{21}\\
			I_7
		\end{smallmatrix}\right),
            \end{split}
	\end{align*}
	\vskip -12pt
	\begin{align*}
           \begin{split}
		\bm r_2
                   &=\left(\begin{smallmatrix}
			u_{11}x_{12}+u_{12}x_{22}\\
			u_{21}x_{12}+u_{22}x_{22}+x_{12}\\
			u_{31}x_{12}+u_{32}x_{22}+x_{22}
		\end{smallmatrix}\right)=\left(\begin{smallmatrix}
			I_2\\
			I_5+x_{12}\\
			I_8+x_{22}
		\end{smallmatrix}\right),
          \end{split}
	\end{align*}
	\vskip -12pt
	\begin{align*}
           \begin{split}
		\bm r_3
                  &=\left(\begin{smallmatrix}
			u_{11}x_{13}+u_{12}x_{23}+x_{23}\\
			u_{21}x_{13}+u_{22}x_{23}\\
			u_{31}x_{13}+u_{32}x_{23}+x_{13}
		\end{smallmatrix}\right)=\left(\begin{smallmatrix}
			I_3+x_{23}\\
			I_6\\
			I_9+x_{13}
		\end{smallmatrix}\right),
              \end{split}
	\end{align*}
	\vskip -12pt
	\begin{align*}
		\bm r_4=\left(\begin{smallmatrix}
			u_{11}&u_{12}\\
			u_{21}&u_{22}\\
			u_{31}&u_{32}
		\end{smallmatrix}\right)\left(\begin{smallmatrix}
			x_{11}+x_{12}\\
			x_{21}+x_{22}
		\end{smallmatrix}\right)=\left(\begin{smallmatrix}
			I_1+I_2\\
			I_4+I_5\\
			I_7+I_8
		\end{smallmatrix}\right),
	\end{align*}
	\vskip -12pt
	\begin{align*}
		\bm r_5=\left(\begin{smallmatrix}
			u_{11}&u_{12}\\
			u_{21}&u_{22}\\
			u_{31}&u_{32}
		\end{smallmatrix}\right)\left(\begin{smallmatrix}
			x_{12}+x_{13}\\
			x_{22}+x_{23}
		\end{smallmatrix}\right)=\left(\begin{smallmatrix}
			I_2+I_3\\
			I_5+I_6\\
			I_8+I_9
		\end{smallmatrix}\right),
	\end{align*}
	where $I_{i}=\sum_{j=1}^2 u_{tj}x_{js}$ and $i=3(t-1)+s$, with $s,t=1,\ldots,3$, are collected by the user. 
 Notice that each storage node sends back $k=3$ symbols. The user obtains the requested file as follows. Knowing $I_2$, the user obtains $I_1$ and $I_3$ from the first components of $\bm r_4$ and $\bm r_5$. This allows the user to obtain $x_{11}$ and $x_{23}$. In a similar fashion, knowing $I_6$ the user gets $I_5$ from the second component of $\bm r_5$, then uses this to obtain $I_4$ from the second component of $\bm r_4$. This allows the user to obtain $x_{21}$ and $x_{12}$. Similarly, knowing $I_7$ allows the user to get $I_8$ from the third component of $\bm r_4$. Knowing $I_8$ allows the user to obtain $I_9$ from the third component of $\bm r_5$, which then allows to recover the symbols $x_{22}$ and $x_{13}$. In this way, the user recovers all symbols of the file and hence recovers $\bm X$. Note that 
	$\theta=\frac{5\cdot3}{2\cdot3}=2.5$,
which is equal to the lower bound $1/(1-R)$.
\end{example}

\section{Optimizing the Communication Price of Privacy}
In the previous section, we provided a construction of a PIR scheme for DSSs that use an arbitrary linear systematic code to store data and showed that a cPoP of $n / (\tilde d_\text{min}-1)$ is achievable while maintaining information-theoretic PIR (see  \cref{th:PIR} and \cref{cor:PIR}). In this section, we provide an algorithm (based on \cref{th:PIR}) to further lower the cPoP taking the structure of the underlying  code into consideration. The algorithm is outlined in \cref{alg:cPoP}.


\setlength{\textfloatsep}{14pt}
\begin{algorithm}[t]
\SetKwFunction{CompEraPat}{ComputeErasurePatternList}
\SetKwFunction{CompEraMat}{ComputeMatrix}
\SetKwInOut{Input}{Input}
\SetKwInOut{Output}{Output}

\Input{Distributed storage code $\mathcal{C}$ of length $n$}
\Output{Optimized matrix $\bm E_{\rm opt}$ and largest possible $\beta$}
$\beta\leftarrow\tilde{d}_{\text{min}}-1$\\
$\bm E_{\rm opt} \leftarrow \emptyset$, $\beta_{\rm opt} \leftarrow \beta$\\
	\While{$\beta\leq \tilde{n}-\tilde{k}$}{\label{algcPoP:outerWhile}
	$\mathcal{L} \leftarrow$ \CompEraPat{$\tilde{\mathcal{C}},\beta$} \label{algcPoP:subprocedure1}\\
		\If{$\mathcal L\not=\emptyset$}{
		$\bm{E} \leftarrow$ \CompEraMat{$\mathcal{L}$} \label{algcPoP:subprocedure2}\\
		\eIf{$\bm{E} \neq \emptyset$}{
		$\bm E_{\rm opt} \leftarrow \bm E$, $\beta_{\rm opt} \leftarrow \beta$\\
		}{
                \KwRet{$(\bm E_{\rm opt}, \beta_{\rm opt})$}
		}
		}
                
		$\beta\leftarrow\beta+1$\\
	}
        \KwRet{$(\bm E_{\rm opt},  \beta_{\rm opt})$}
	\caption{Optimizing the cPoP}\label{alg:cPoP}
\end{algorithm}

The main issues that need to be addressed are the efficient enumeration of the set of erasure patterns of a given weight $\beta$ that can be corrected under ML decoding of $\tilde{\mathcal{C}}$, and the efficient computation of the matrix $\bm E$. Such ML-correctable erasure patterns are binary vectors (of length $\tilde{n}=k$) which can be ML-decoded on the BEC when the positions corresponding to the $1$-entries are erased by the channel.  These issues are addressed by the subprocedures \texttt{ComputeErasurePatternList}($\tilde{\mathcal{C}},\beta$) and \texttt{ComputeMatrix}($\mathcal{L}$), in \cref{algcPoP:subprocedure1,algcPoP:subprocedure2}, respectively, of \cref{alg:cPoP}, 
and discussed below in \cref{sec:ComputeList,sec:computeE}.

We remark that the algorithm will always return $\bm E_{\rm opt} \neq \emptyset$, since initially $\beta = \tilde{d}_{\rm min}-1$. Then, in the first iteration of the main loop, the list $\mathcal{L}$ will contain all length-$k$ binary vectors of weight $\beta= \tilde{d}_{\rm min}-1 < \tilde{d}_{\rm min}$. Thus, any vector that is shift-variant (i.e., the $k$ cyclic shifts are all different) can be chosen for the first row of $\bm E$. The remaining rows of $\bm E$ are obtained by cyclically shifting the first row (the $i$-th row is obtained by cyclically shifting the first row $i$ times). In the particular case of $\mathcal{C}$ being  an MDS code, $d_{\rm min}=\tilde{d}_{\rm min}=n-k+1$, the algorithm will do exactly one iteration of the main loop, and the overall PIR scheme reduces to the one described in \cite[Sec.~IV]{taj16}.


\subsection{\texttt{ComputeErasurePatternList}($\tilde{\mathcal{C}},\beta$)} \label{sec:ComputeList}


Computing a list of erasure patterns that are correctable under ML decoding for a given short code can be done using any ML decoding algorithm. For small codes $\tilde{\mathcal{C}}$, all length-$k$ binary vectors of weight $\beta$ that are ML-correctable can be found using an exhaustive search, while for longer codes a random search can be performed, in the sense of picking length-$k$ binary vectors of weight $\beta$ at random, and then verifying whether or not they are ML-correctable. Alternatively, one can apply a random permutation $\pi$ to the columns of $\tilde{\bm H}$, apply the Gauss-Jordan algorithm on the resulting matrix to transform it into \emph{row echelon form}, collect a subset of size $\beta$ of the column indices of \emph{leading-one-columns}, and finally apply the inverse permutation of $\pi$ to this subset of column indices. The leading-one-columns are the columns containing a \emph{leading one}, where the first nonzero entry in each matrix row of a matrix in row echelon form is called a leading one.
%
%
%
%
%
 This will give the support set of an ML-correctable erasure pattern of weight $\beta$ that can be added to $\mathcal{L}$. Finally, one can check whether all cyclic shifts of the added erasure pattern are ML-correctable or not and add the ML-correctable cyclic shifts to $\mathcal{L}$.  

\subsection{\texttt{ComputeMatrix}($\mathcal{L}$)} \label{sec:computeE}

Given the list $\mathcal{L}$ of erasure patterns that are correctable under ML decoding, we construct a $|\mathcal{L}| \times k$ matrix, denoted by $\bm \Psi = [\psi_{ij}]$,  in which each row is one of these patterns. The problem is now to find a $k \times k$ submatrix of constant column weight $\beta$ (and constant row weight $\beta$). This can be formulated as an integer program (in the integer variables $\eta_1,\eta_2,\ldots,\eta_{|\mathcal{L}|}$) in the following way,
\begin{equation} \label{eq:LIP}
\begin{array}{rl}
\text{maximize} &  \sum_{i=1}^{|\mathcal{L}|} \eta_i \\
\text{s.\,t.} & \sum_{i=1}^{|\mathcal{L}|} \eta_i \psi_{ij} = \beta,\, \forall j \in \{1,\ldots,k\}, \\ 
& \eta_i\in\{0,1\},\, \forall i\in\{1,\ldots,|\mathcal{L}|\}, \text{ and } \\
& \sum_{i=1}^{|\mathcal{L}|} \eta_i = k.
\end{array}
\end{equation}
A valid $k \times k$ submatrix of $\bm \Psi$ is constructed from the rows of $\bm \Psi$ with corresponding $\eta$-values of one in any feasible solution of (\ref{eq:LIP}). When $|\mathcal{L}|$ is large, solving  (\ref{eq:LIP}) may become impractical (solving a general integer program is known to be NP-hard), in which case one can take several random subsets of the list $\mathcal{L}$ of some size, construct the corresponding matrices $\bm \Psi$, and try to solve the program in (\ref{eq:LIP}). Finally, before solving (\ref{eq:LIP}), one may check whether there are erasure patterns in $\mathcal{L}$ with all its cyclic shifts (assuming they are all different) also in   $\mathcal{L}$, in which case the corresponding submatrix of $\bm \Psi$ will be a valid $k \times k$ matrix $\bm E$.


\section{Numerical Results}

We present optimized values for the cPoP for different systematic linear codes. The results are tabulated in \cref{table_of_codes}, where $\theta_{\rm LB} = 1/(1-R)$ is the lower bound on the cPoP taken from \cite[Th.~5]{cha15},  $\theta_{\rm opt}$ is the optimized value computed from \cref{alg:cPoP}, and $\theta_{\rm non-opt} = n/(\tilde d_\text{min}-1)$. The code $\mathcal C_1$ in the table is from \cref{ex:5_3_code}, $\mathcal C_2$ is an $(11,6)$ binary linear code with optimum minimum distance, while codes $\mathcal C_3$ and $\mathcal C_5$ are Pyramid codes, taken from \cite{Hua07}, of locality of $4$ and $6$, respectively. $\mathcal C_4$ is an LRC of locality $5$ borrowed from \cite{Sat13}. 

In \cite{hao16}, a construction of optimal (in terms of minimum distance) binary LRCs with multiple repair groups was given. In particular, in \cite[Constr.~3]{hao16}, a construction based on array low-density parity-check (LDPC) codes was provided. The parity part (or the $\bm P$ matrix)  of the parity-check matrix $\bm H$ of the optimal LRC is the parity-check matrix of an array-based LDPC code. The minimum distance of array LDPC codes is known for certain sets of parameters (see, e.g., \cite{ros14}, and references therein). Codes $\mathcal{C}_6$ and $\mathcal{C}_7$ in \cref{table_of_codes} are \emph{optimal} LRCs based on array LDPC codes constructed using \cite[Constr.~3]{hao16} and having a locality of $11$.

For all codes, $\theta_{\rm opt}$ is close to the lower bound on the cPoP, $\theta_{\rm LB}$. Remarkably, the codes $\mathcal C_1$, $\mathcal C_3$, and $\mathcal C_5$ achieve the lower bound on the cPoP despite the fact that these codes are not MDS codes. The remaining codes ($\mathcal C_2$,  $\mathcal C_4$, $\mathcal C_6$, and $\mathcal C_7$) achieve a cPoP of $\tilde{n}/(\tilde{n} - \tilde{k}) = k/(k-\tilde{k}) > 1/(1-R)$, which is the lowest possible value given the parameters of $\tilde{\mathcal{C}}$. The strict inequality is due to the fact that  $\tilde{\bm H}$ is not full rank. 

\setlength{\textfloatsep}{18.60004pt plus 2.39996pt minus 4.79993pt}
\begin{table}[!t]
     \caption{Optimized values for the cPoP for different codes}
    \label{table_of_codes}
    \centering
    \def\Hline{\noalign{\hrule height 2\arrayrulewidth}}
    \vskip -2.0ex 
    \begin{tabular}{@{}lccccc@{}}
        \Hline \\ [-2.0ex]
        Code & $d_{\rm min}$ & $\tilde{d}_{\rm min}$ & $\theta_{\rm non-opt}$ & $\theta_{\rm opt}$ & $\theta_{\rm LB}$ \\
        \hline
        \\ [-2.0ex] \hline  \\ [-2.0ex]
        $\mathcal C_1:(5,3)$ (\cref{ex:5_3_code}) & $2$ & $3$ &  $2.5$ & $2.5$ & $2.5$ \\
        $\mathcal C_2:(11,6)$  & $4$ & $4$ & $3.6667$ & $2.75$ & $2.2$ \\
        $\mathcal C_3:(12,8)$ Pyramid & $4$ & $4$ & $4$ & $3$ & $3$ \\
        $\mathcal C_4:(16,10)$ LRC & $5$ & $5$ & $4$ & $3.2$ & $2.6667$ \\
        $\mathcal C_5:(18,12)$ Pyramid & $5$ & $5$ & $4.5$ & $3$ & $3$ \\
        $\mathcal C_6:(154,121)$ LRC & $4$ & $6$ & $30.8$ & $4.9677$ & $4.6667$ \\
        $\mathcal C_7:(187,121)$ LRC & $7$ & $16$ & $12.4667$ & $3.0656$ & $2.8333$ \\
        \hline
    \end{tabular}
    \vspace{-0.5ex}
    \vskip -2ex 
\end{table}

\section{Conclusion}
We generalized the PIR protocol proposed in \cite{taj16} for a DSS with a single spy node and where data is stored using an MDS code to the case where an arbitrary systematic linear code of rate $R>1/2$ is used to store data. We also presented an algorithm to optimize the cPoP of the protocol. The optimization leads to a cPoP close to its theoretical lower bound. Interestingly, for certain codes, the lower bound on the cPoP can be achieved.

\appendices
\section*{Appendix\\Proof of \cref{th:PIR}}
\label{Appendix: Proof}
	Consider the $t$-th subresponse of each response $\bm r_j$. Out of the $k$ subresponses generated from the systematic nodes, there are $\beta$ subresponses originating from a subset of systematic nodes $\mathcal J\subset\{1,\ldots,k\}, |\mathcal J|=\beta$, of the form 
	\begin{align}
		\label{Eq: P1E1}
		I_{(t-1)k+j_1}+x^{(m)}_{a,j_1}=r_{j_1,t},
	\end{align}
	where $j_1\in\mathcal J$, $a\in\{1,\ldots,\beta\}$, and  
	 $I_{(t-1)k+j_1}=\sum_{m=1}^{f}\sum_{j=(m-1)\beta+1}^{m\beta} u_{t,j}x_{j-(m-1)\beta,j_1}^{(m)}$ is the interference symbol. 
	 The subresponses from the remaining $k-\beta$ systematic nodes in $\{1,\ldots,k\}\setminus\mathcal J$ are
	\begin{align}
		\label{Eq: P1E2}
		I_{(t-1)k+j_2}=r_{j_2,t},
	\end{align}
	where $j_2\in\{1,\ldots,k\}\setminus\mathcal J$ and 
	$I_{(t-1)k+j_2}=\sum_{m=1}^{f}\sum_{j=(m-1)\beta+1}^{m\beta} u_{t,j}x_{j-(m-1)\beta,j_2}^{(m)}$ 
	is the interference symbol. It is trivial to see that \cref{Eq: P1E1} and \cref{Eq: P1E2} result in a system of linear equations that is underdetermined in the unknowns $I_{(t-1)k+j_1}$, $I_{(t-1)k+j_2}$, and $x_{a,j_1}^{(m)}$. In order to retrieve the $\beta$ message symbols, we require the knowledge of the interference symbols $I_{(t-1)k+j_1}$ (see \cref{Eq: P1E1}). The subresponses from the parity nodes are given by
	\begin{align}
		\label{Eq: P1E3}
		r_{j,t}=
				 \sum_{i=1}^{k} \lambda_{j,i}I_{(t-1)k+i},
	\end{align}
where $j=k+1,\ldots,n$ and $\lambda_{j,i}\in\text{GF}(q^\alpha)$ is the coefficient associated to the message symbol in the $i$-th node that is used in the weighted sum of the code (parity) symbol in the $j$-th node. $I_{(t-1)k+i}$ is an interference symbol used in either \cref{Eq: P1E1} or \cref{Eq: P1E2}. We can interpret \cref{Eq: P1E3} as a parity-check equation of the $(\tilde{n}=k,\tilde{k} \geq 2k-n)$ code $\tilde{\mathcal{C}}$, where the interference symbols $I_{(t-1)k+i}$ form the message symbols. Since the $k-\beta$ interference symbols $I_{(t-1)k+j_2}$ from (\ref{Eq: P1E2}) are known, \cref{Eq: P1E3} reduces to
	\begin{align}
		\label{Eq: P1E4}
		\tilde{r}_{j,t}= \sum_{i\in\mathcal J} \lambda_{j,i}I_{(t-1)k+i}.
	\end{align}
	Solving the system of linear equations in \cref{Eq: P1E4} in the unknowns $I_{(t-1)k+i}$, $i \in \mathcal J$, is now just a decoding problem of the aforementioned code. 
Since there are $\beta$ unknowns and \cref{Eq: P1E4} corresponds to an ML-corretable erasure pattern (from the third requirement for $\bm E$ in \cref{sec:constr}), \cref{Eq: P1E4} becomes a full-rank linear system of equations in $\text{GF}(q^{\ell\alpha})$. Hence, knowing the interference symbols allows the recovery of $\beta$ \emph{unique} (from the first requirement for $\bm E$ in \cref{sec:constr}) message symbols in the $t$-th subquery. 
Combined with the second requirement for $\bm E$ in \cref{sec:constr}, it follows that all $k$ subqueries yield $k\beta$ \emph{unique}  message symbols of the requested file $\bm X^{(m)}$, from which it follows that $H(\bm X^{(m)}|\bm r_1,\bm r_2,\ldots, \bm r_n)=0$.

\balance



\end{document}